\documentclass[referee]{aa} 
\usepackage{graphicx}
\begin{document}
\title{Optical photometry and spectral classification in the field
      of the open cluster NGC~6996 in the North America Nebula
       \thanks{{}Based on observations carried out at Asiago and  Teramo
Observatories, Italy},
       \thanks{Photometry is only available in electronic form at the CDS 
               via anonymous ftp to {\tt cdsarc.u-strasbg.fr (130.79.128.5)} or 
	       via {\tt http://cdsweb.u-strasbg.fr/cgi-bin/qcat?J/A+A//}
	       }
       }
       \author{{}Sandro Villanova\inst{1},
             Gustavo Baume\inst{1,2},
		 Giovanni Carraro\inst{1}
             \and
                Anna Geminale\inst{1}
		 }

   \offprints{villanova@pd.astro.it}

   \institute{Dipartimento di Astronomia, Universit\`a di Padova,
              Vicolo Osservatorio 2, I-35122 Padova, Italy
         \and
             Facultad de Ciencias Astron\'omicas y Geof\'{\i}sicas de la
	     UNLP, IALP-CONICET, Paseo del Bosque s/n, La Plata, Argentina
	     }

   \date{Received **; accepted **}

\abstract{
We present and discuss
broad band CCD $UBV(I)_C$ photometry and low resolution spectroscopy 
for stars in the region of the open cluster NGC~6996, located in the North
America Nebula. The new data allow us to tightly constrain the 
basic properties of this object. We revise the cluster
size, which in the past has been significantly underestimated.
The width of the Main Sequence is mainly
interpreted in terms of differential reddening, and indeed the stars'
color excess $E_{B-V}$ ranges from $0.43$ to $0.65$, implying the presence
of a significant and evenly distributed dust component.
We cross-correlate our optical photometry with near infrared from 2MASS, 
and by means of spectral classification we are able to build up 
extinction curves for an handful of bright members. We find that the 
reddening slope and the total to selective absorption ratio $R_V$
toward NGC~6996 are anomalous. Moreover 
the reddening corrected colors and magnitudes allow us to derive
estimates for the cluster distance and age, which turn out to be
$760 \pm 70 pc$ ($V_{0}-M_{V} = 9.4 \pm 0.2$) and $\sim 350$ Myr,
respectively. Basing on our results, we suggest that NGC~6996 is located in
front of the North America Nebula, and does not seem to have any apparent 
relationship with it.
\keywords{Galaxy: open clusters and associations: individual: NGC 6996 
-- open clusters and associations: general}
}

\authorrunning{Villanova et al.}
\titlerunning{The open cluster NGC~6996}

\maketitle
%

\section{Introduction}

$\hspace{0.5cm}$
Galactic open clusters are fundamental tools to probe 
the global properties and evolution of the Galactic disk. 
The determination of their fundamental parameters 
(age, distance, reddening pattern, size) allows us 
not only to better understand the properties and evolution
of the Galactic open cluster system as a whole, but in many cases to put
constrains on large
structures like super-clusters, complexes, 
H~II regions and molecular clouds systems, which they might be 
part of or not. \\
This is the case of NGC~6996 = C2054+444 
($l = 85.47^{\circ}$, $b = -0.48^{\circ}$), which is   
believed to lie
in front of  the North America Nebula (NGC~7000) HII region, 
precisely near its western edge (see Fig.~1), where 
an important dust cloud separates this  nebula from the nearby
Pelican Nebula. \\
NGC~6996 is a sparse and moderately young open cluster ($\sim 10^{8}$ Myr, 
Zdanavicius \& Straizys 1990). 
However, its basic parameters, in particular  distance and  reddening,
are not much well known, and this was the main motivation driving
this study.\\
Previous investigations carried out in this area include the photographic 
studies by Muller (1936) and Barkhatova (1958). Later, a more detailed
study has been conducted by Zdanavicius \& Straizys (1990),
who  obtained photoelectric measurements of several 
bright stars in the region of NGC~6996 in the Vilnius system. 
More recently, 
Subramaniam et al. (1995) catalogued this cluster as a probable 
binary one together with the twin 
cluster Collinder~428. This latter 
object is located in the opposite (the eastern) border of the 
North America Nebula, and indeed both clusters might be 
related with the H~II region. The entire 
area belongs to the Cygnus region, which has been studied rather intensively. 

\begin{figure*}
\centering
\includegraphics[width=14cm,height=10cm]{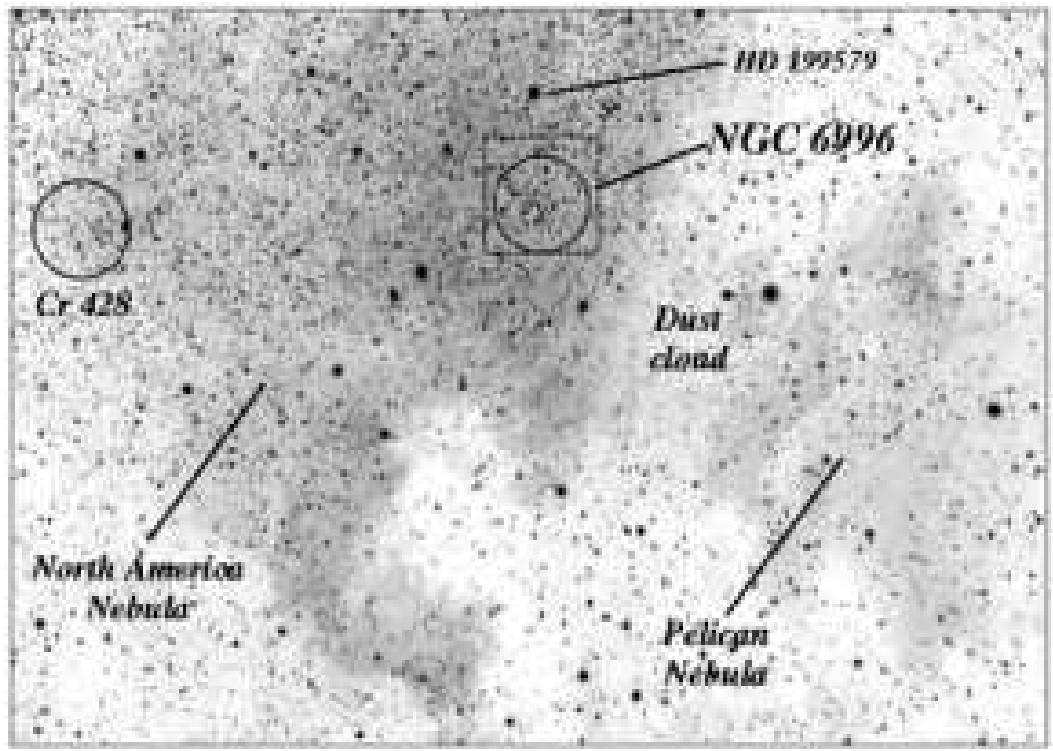}
\caption{A map of the North America Nebula region. The region of NGC~6996 covered
by our study is marked by a dashed square (see also Fig.~3). North is up,
East on the left.}
\end{figure*}
\noindent
In fact this region (from $l \sim 74^{\circ}$ to $l \sim 85^{\circ}$)
has the singular property that stars present an 
anomalous, with a larger slope, interstellar reddening law 
(see Straizys et al. 1999 and references therein). 
In particular, the visual absorption distribution 
in this region was studied by Goudis \& White (1979) with the $H_{\alpha}$ surface  brightness technique, by Bally \& Scoville (1980) 
by using $^{12}$CO observations and, finally, by 
Cambr\'{e}sy et al. (2002) from 2MASS data. These latter authors by the way discovered some new possible star
clusters toward this area, which might be  physical objects and as a 
consequence deserve further investigation.\\
In an attempt to derive better estimates of the NGC~6996 basic parameters 
(reddening, distance and age), to study the reddening law in 
this region and to understand whether some
relations exist between the cluster and the North America Nebula, we performed 
 CCD $UBV(I)_{C}$ photometry covering most of the cluster region,
and complement it with
low resolution spectroscopic observation of bright stars in this region. 
Our data were also 
cross-correlated with near infrared ($JHK_S$) data from the 2MASS catalogue and with 
astrometric information available from the Tycho-2 catalogue 
(H{\o}g et al. 2000) for the brightest stars. \\

The layout of the paper is as follows.
In Sect.~2 we describe our observations and the reduction procedure. In Sect.~3
we present the data analysis. Sect.~4 deals with  the cluster
basic parameters determination. 
Sect.~5 is finally devoted to a brief discussion of the  
outcomes of this study.

\begin{table} 
\fontsize{8} {10pt}\selectfont
\caption{Journal of observations of NGC 6996 and standard star fields together 
with calibration coefficients.} 
\begin{tabular}{ccccccc} 
\hline 
\multicolumn{1}{c}{Field}         & 
\multicolumn{1}{c}{Filter}        & 
\multicolumn{3}{c}{Exposure time} & 
\multicolumn{1}{c}{Seeing}        &
\multicolumn{1}{c}{Air-mass}       \\
 & & \multicolumn{3}{c}{[sec.]} & [$\prime\prime$] & \\ 
\hline 
NGC 6996          & U &  900x2 & 180 & 10 & 2.2 & 1.091 \\
1 frame           & B &  600   &  60 & 10 & 2.1 & 1.062 \\
Asiago            & V &  300   &  30 &  5 & 2.2 & 1.052 \\
8-11-02           & I &  300   &  30 &  5 & 2.3 & 1.045 \\
[1 ex]
NGC 6996          & U &  900   &   - &  - & $\sim 3$ & 1.368 \\
2 frames          & B &  600   &   - &  - & $\sim 3$ & 1.322 \\
Teramo            & V &  300   &  60 &  - & $\sim 3$ & 1.251 \\
5-12-2002          & I &  180   &  60 &  - & $\sim 3$ & 1.219 \\
[1 ex]
NGC 6996          & U &  1200  & 120 &  - & 1.9 & 1.007 \\
1 frame           & B &  800   &  80 &  8 & 1.8 & 1.018 \\
Asiago            & V &  600   &  60 &  6 & 1.8 & 1.027 \\
18-09-03          & I &  400   &  40 &  4 & 1.7 & 1.037 \\
\hline 
PG 0231+051       & U &  800   &     &    & 2.5 & 1.348 \\
Asiago            & B &  300   &     &    & 2.4 & 1.324 \\ 
8-11-02           & V &   60   &     &    & 2.2 & 1.316 \\ 
                  & I &   90   &     &    & 2.2 & 1.315 \\ 
[1 ex]
PG 2213-006       & U &  600   &     &    & 2.5 & 1.447 \\
Asiago            & B &  150   &     &    & 2.3 & 1.457 \\ 
8-11-02           & V &   30   &     &    & 2.3 & 1.465 \\ 
                  & I &   30   &     &    & 2.3 & 1.472 \\ 
\hline
\hline
Calibration       & \multicolumn {3}{l}{$u_1 = +3.861 \pm 0.015$}     & \multicolumn {3}{l}{$b_1 = +1.602 \pm 0.004$} \\
coefficients      & \multicolumn {3}{l}{$u_2 = -0.142 \pm 0.022$}     & \multicolumn {3}{l}{$b_2 = +0.038 \pm 0.006$} \\
Asiago            & \multicolumn {3}{l}{$u_3 = +0.58$}                & \multicolumn {3}{l}{$b_3 = +0.29$}            \\
8-11-02		& \multicolumn {3}{l}{$v_{1bv} = +1.003 \pm 0.014$} & \multicolumn {3}{l}{$i_1 = +1.691 \pm 0.044$} \\
                  & \multicolumn {3}{l}{$v_{2bv} = -0.016 \pm 0.018$} & \multicolumn {3}{l}{$i_2 = +0.057 \pm 0.043$} \\
                  & \multicolumn {3}{l}{$v_3 = +0.16$}                & \multicolumn {3}{l}{$i_3 = +0.08$}            \\
                  & \multicolumn {3}{l}{$v_{1vi} = +1.002 \pm 0.016$} & \\
                  & \multicolumn {3}{l}{$v_{2vi} = -0.013 \pm 0.016$} & \\
\hline
\end{tabular}
\end{table}
\begin{figure}
\centering
\includegraphics[width=7.5cm]{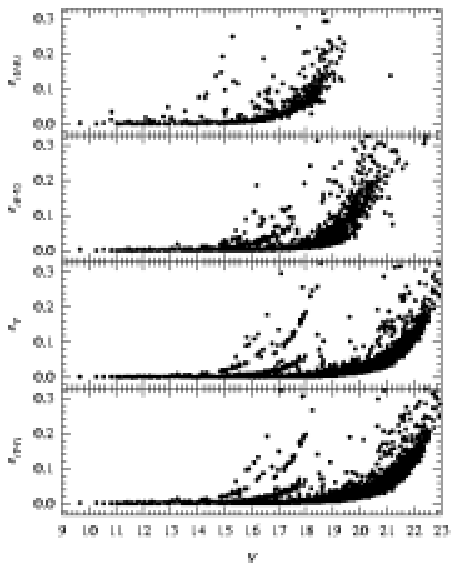}
\caption{DAOPHOT errors in the colour indexes and $V$ magnitude as a function of $V$}
\end{figure}

\section{Observations and data reduction}

\subsection{Photometry}

$\hspace{0.5cm}$
CCD $UBV(I)_C$ data were obtained during three observational runs. Two of them 
were carried out with the AFOSC camera at the 1.82m Copernico telescope of Cima 
Ekar (Asiago, Italy), in the photometric nights of November 8, 2002 and September 
18, 2003. AFOSC samples a $8\farcm14 \times 8\farcm14$ field in a 
$1K \times 1K$ nitrogen-cooled thinned CCD and this camera was used to take a 
central and a south frame in the cluster area. The other run was conducted at 
Teramo Observatory (Italy) using a $512 \times 512$ CCD onboard the 0.72m 
Teramo Normale Telescope (TNT). Teramo observations were used to complement Asiago ones 
since two peripheral and less deep field were taken during a non photometric 
night. Figs.~1 and 3 show the finding charts of the covered area. \\

\renewcommand{\thefootnote}{\dag}
Details of the observations are listed in Table~1, where the observed fields are
reported together with the exposure times, the typical seeing values and the 
air-masses. The data has been reduced with the
IRAF\footnote{IRAF is distributed by NOAO, which are operated by AURA under 
cooperative agreement with the NSF.} 
packages CCDRED, DAOPHOT, and PHOTCAL using the point spread function (PSF)
method (Stetson 1987) for the frame obtained on 8/11/02 (Asiago) and using only 
aperture photometry for the others. The calibration equations obtained by 
observing Landolt (1992) PG 0231+051 and PG 2213-006 fields at the Asiago 
Observatory, are: 

\begin{center}
\begin{tabular}{lc}
$u = U + u_1 + u_2 (U-B) + u_3 X$         & (1) \\
$b = B + b_1 + b_2 (B-V) + b_3 X$         & (2) \\  
$v = V + v_{1bv} + v_{2bv} (B-V) + v_3 X$ & (3) \\  
$v = V + v_{1vi} + v_{2vi} (V-I) + v_3 X$ & (4) \\  
$i = I + i_1 + i_2 (V-I) + i_3 X$         & (5) \\
\end{tabular}
\end{center}

$\hspace{-0.7cm}$
where $UBVI$ are standard magnitudes, $ubvi$ are the instrumental ones, $X$ is 
the air-mass and the derived coefficients are presented in the bottom of Table~1.
Since observations from Teramo Observatory were taken during a non
photometric night, they were zero-point and color-term related to the Asiago data. 
As for $V$ magnitudes, when $B$ magnitude was 
available, we used expression (3) to compute them, elsewhere expression (4) was 
used. The standard stars in these fields provide a very good color coverage which 
allows us to obtain reliables transformations. For the extinction coefficients, we 
assumed the typical values for the Asiago Observatory (Desidera et al. 2002
\footnote{www.pd.astro.it/Asiago/5000/5100/5100.html}). 
The photometric error trends against the $V$ magnitude data are shown in 
Fig.~2, where one can clearly distinguish the error trends for TNT and Asiago
observations.

The photometric data for some of the brightest stars in the region of 
NGC~6996 are shown in Table~2. The full table is only available electronically
at CDS. \\

\begin{figure*}
\centering
\includegraphics[width=12cm]{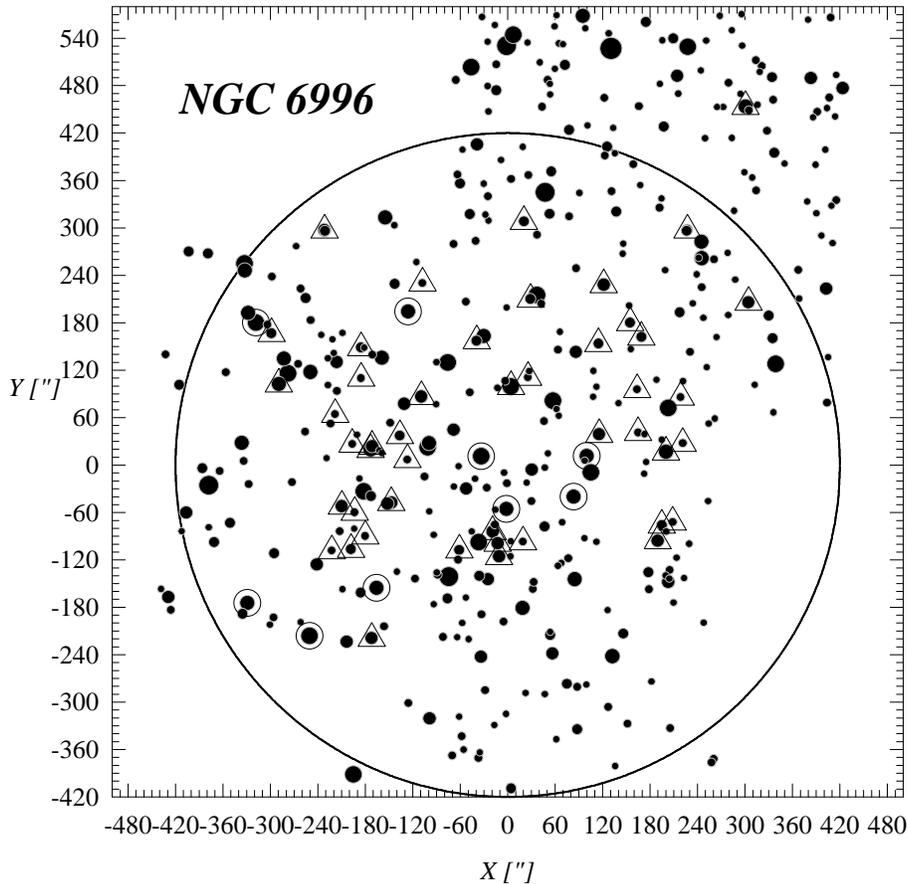}
\caption{Finding chart of the NGC 6996 region ($V$ filter). The black solid circle 
indicates the adopted angular size (radius =$7^{\prime}$) for the cluster (see
Sect~3.1 and Fig.~5). 
Adopted likely and probable cluster members are enclosed in small circles and
triangles, respectively. For a coordinate reference, the center ($X = 0$; $Y=0$) 
corresponds to the cluster coordinates (see Sect~3.1) and each $X$-$Y$ are 
expressed in arcsecs. North is up, East on the left.}
\end{figure*}

\begin{figure*}
\centering
\includegraphics[width=14cm]{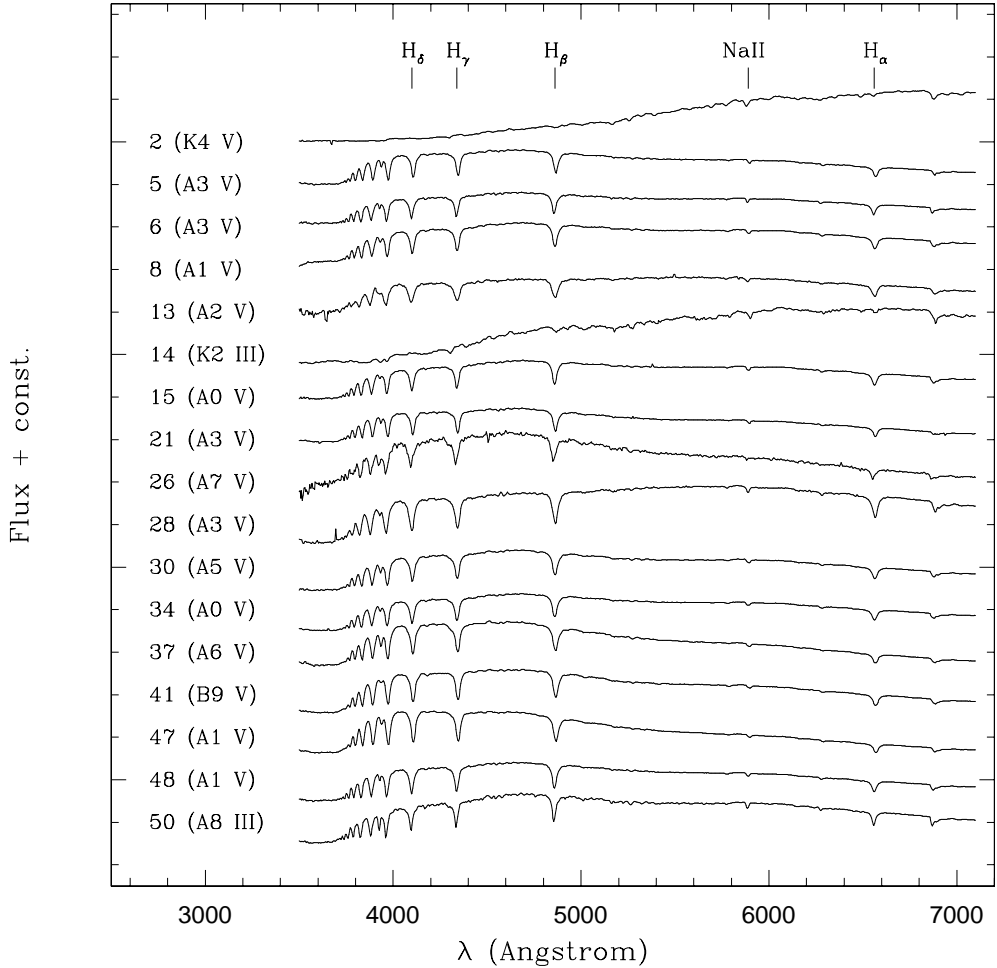}
\caption{Spectra of some bright stars in the field of NGC~6996. A few 
interesting lines are indicated. See Table~2 for details. Numbers correspond 
to Zdanavicius \& Straizys (1990) identification.}
\end{figure*}

\subsection{Spectroscopy}
$\hspace{0.5cm}$
Spectroscopic observations were carried out with the AFOSC camera at the 1.82m 
Copernico telescope of Cima Ekar (Asiago, Italy), in the following nights:
July 21, September 18 and  October 1, 2003.
The instrument was used in low dispersion mode (R=600), the grism $\#4$ 
was chosen to have a large spectral coverage (3500-7000 A) and the exposure 
time ranged from 10 to 30 minutes,  according to the brightness of the stars.
The data has been reduced with the IRAF package for 
one-dimensional spectroscopy {\em CTIOSLIT} and by using the Hg-Cd lamp spectra 
for wavelength calibration purposes.
In order to have flux-calibration, a few standard spectrophotometric
stars were also  observed.
Spectra (see Fig.~4) were classified in two different ways:
\begin{itemize}
\item by comparing them with a library of spectroscopic standards
(e.g. Jacoby et al. 1984; Torres-Dodgen \& Weawer 1993);
\item by measuring the equivalent width of some absorption lines 
(mainly $H_{\alpha}$, $H_{\beta}$ and  $H_{\gamma}$) (see Jaschek \& Jaschek 1987).
\end{itemize}
\noindent
At this resolution we estimate an error in the spectral type derivation of $\pm$1
tenth.  
The derived spectral types and luminosity classes are reported in Table~2
and Fig.~4.

\section{Data Analysis}

\subsection{Cluster angular size}
$\hspace{0.5cm}$

\renewcommand{\thefootnote}{\ddag}
To derive an estimate of the cluster angular size, we computed surface stellar
densities at increasing concentric $1^{\prime}$ wide annuli around the 
adopted cluster center ($\alpha_{2000} = 20:56:30$; $\delta_{2000} = +44:38:00$)
over: $a)$ the corresponding DSS-2\footnote{Second generation Digitized Sky Survey, 
{\tt http://cadcwww.dao.nrc.ca/cadcbin/getdss}} 
red image, and $b)$ the 2MASS infrared data. The results are shown in Fig.~5,
where the dotted line is the field star density level as derived 
from mean star counts in two 2MASS
fields, 20$^{\prime}$ southward and northward NGC~6996, respectively.
The patchy distribution of dust and gas clearly renders it difficult
to fix a cluster radius. However,
by inspecting both DSS maps and this plot we argue that 
the cluster radius is 
$\sim 7^{\prime}$, the point (dashed vertical line in Fig.~5) at which the 
stellar density more clearly reaches the field 
level. 
Therefore, we adopt this value as  angular radius, which turns out
to be  almost twice the estimates reported  by Lyng\aa (1987) and Dias et al. (2002) 
($6^{\prime}$ and $8^{\prime}$ in diameter, respectively). \\

\begin{table*}
\fontsize{8} {10pt}\selectfont
\caption{{} Brightest stars in the region of NGC 6996.}
\begin{center}
\begin{tabular}{rrlrrcccccc}
\hline
  \# & \#$_{ZS}$ & 2MASS ID.         & $X~[^{\prime\prime}]$ & $\alpha_{2000}$ & $V$ & $B-V$ & $E_{B-V}$ & $\mu_{\alpha} \cos(\delta)~[mas/yr]$ & $ST$ & Memb. \\
     &           & Tycho-2 ID.       & $Y~[^{\prime\prime}]$ & $\delta_{2000}$ &     & $U-B$ & $E_{U-B}$ & $\mu_{\delta}~[mas/yr]$              &      &       \\
     &           & BD ID.            &                       &                 &     & $V-I$ & $E_{V-I}$ &                                      &      &       \\
\hline
   2 &        26 & J20563677+443538  &  -74.6 & 20:56:36.9 & 10.27 &  0.35 &  0.15 &  4.1 $\pm$ 1.4 & A7 V   &  $nm$ \\
     &           & TYC 3179-687-1    & -141.5 & 44:35:37.9 &       & -0.08 & -0.18 &  6.1 $\pm$ 1.4 &        &       \\
     &           & BD +44 3638       &        &            &       &  0.24 &       &                &        &       \\
[1 ex]
  13 &    50$^*$ & J20570129+444213  & -332.6 & 20:57:01.2 & 11.44 &  0.64 &  0.39 & -2.8 $\pm$ 2.3 & A8 III &  $nm$ \\
     &           & TYC 3179-658-1    &  254.7 & 44:42:13.2 &       &  0.40 &  0.30 & -8.6 $\pm$ 2.2 &        &       \\
     &           &                   &        &            &       &  0.89 &       &                &        &       \\
[1 ex]
  14 &        47 & J20565989+444058  & -318.3 & 20:56:59.9 & 11.48 &  0.45 &  0.44 & -3.5 $\pm$ 2.3 & A1 V   &  $lm$ \\
     &           & TYC 3179-805-1    &  179.9 & 44:40:58.3 &       &  0.35 &  0.33 & -7.8 $\pm$ 2.3 &        &       \\
     &           &                   &        &            &       &  0.66 &  0.64 &                &        &       \\
[1 ex]
  15 &        14 & J20562963+443939  &    4.1 & 20:56:29.6 & 11.56 &  1.44 &  0.52 &                & K2 III &  $pm$ \\
     &           &                   &   99.1 & 44:39:39.2 &       &  1.33 &  0.44 &                &        &       \\
     &           &                   &        &            &       &  1.74 &       &                &        &       \\
[1 ex]
  17 &        21 & J20563310+443810  &  -33.4 & 20:56:33.1 & 11.60 &  0.52 &  0.44 & -3.7 $\pm$ 2.3 & A3 V   &  $lm$ \\
     &           & TYC 3179-87-1     &   11.0 & 44:38:10.7 &       &  0.45 &  0.37 & -8.9 $\pm$ 2.1 &        &       \\
     &           &                   &        &            &       &  0.81 &  0.72 &                &        &       \\
[1 ex]
  23 &        41 & J20565330+443422  & -250.4 & 20:56:53.3 & 12.02 &  0.51 &  0.58 &                & B9 V   &  $lm$ \\
     &           &                   & -216.5 & 44:34:21.9 &       &  0.33 &  0.53 &                &        &       \\
     &           &                   &        &            &       &  0.83 &  0.89 &                &        &       \\
[1 ex]
  24 &    37$^*$ & J20564701+443725  & -182.1 & 20:56:47.0 & 12.02 &  0.41 &  0.24 &                & A6 V   &  $nm$ \\
     &           &                   &  -33.6 & 44:37:25.4 &       &  0.32 &  0.22 &                &        &       \\
     &           &                   &        &            &       &  0.57 &       &                &        &       \\
[1 ex]
  27 &        34 & J20564536+443523  & -165.8 & 20:56:45.4 & 12.10 &  0.57 &  0.59 &                & A0 V   &  $lm$ \\
     &           &                   & -155.6 & 44:35:23.4 &       &  0.41 &  0.43 &                &        &       \\
     &           &                   &        &            &       &  1.05 &  1.06 &                &        &       \\
[1 ex]
  28 &    48$^*$ & J20570072+443503  & -329.3 & 20:57:00.7 & 12.25 &  0.53 &  0.52 &                & A1 V   &  $lm$ \\
     &           &                   & -174.7 & 44:35:03.4 &       &  0.44 &  0.42 &                &        &       \\
     &           &                   &        &            &       &  0.86 &  0.84 &                &        &       \\
[1 ex]
  29 &         5 & J20562065+443811  &   99.9 & 20:56:20.6 & 12.25 &  0.56 &  0.48 & -1.4 $\pm$ 2.0 & A3 V   &  $lm$ \\
     &           & TYC 3179-81-1     &   11.4 & 44:38:11.9 &       &  0.49 &  0.41 & -8.4 $\pm$ 1.9 &        &       \\
     &           &                   &        &            &       &  0.95 &  0.86 &                &        &       \\
[1 ex]
  31 &         2 & J20561119+443817  &  200.2 & 20:56:11.2 & 12.31 &  1.42 &  0.38 &              & K4 V   &  $nm$ \\
     &           &                   &   16.4 & 44:38:17.4 &       &  2.94 &       &              &        &       \\
     &           &                   &        &            &       &  2.40 &       &              &        &       \\
[1 ex]
  32 &        15 & J20563011+443704  &   -1.8 & 20:56:30.1 & 12.34 &  0.57 &  0.59 &              & A0 V   &  $lm$ \\
     &           &                   &  -55.6 & 44:37:04.3 &       &  0.46 &  0.48 &              &        &       \\
     &           &                   &        &            &       &  0.97 &  0.98 &              &        &       \\
[1 ex]
  34 &         8 & J20562200+443537  &   84.8 & 20:56:21.9 & 12.52 &  0.73 &  0.72 &              & A1 V   &  $nm$ \\
     &           &                   & -144.7 & 44:35:35.5 &       &  0.38 &  0.36 &              &        &       \\
     &           &                   &        &            &       &  0.56 &       &              &        &       \\
[1 ex]
  36 &        30 & J20564188+444113  & -125.9 & 20:56:41.9 & 12.56 &  0.63 &  0.48 &              & A5 V   &  $lm$ \\
     &           &                   &  194.1 & 44:41:13.5 &       &  0.45 &  0.35 &              &        &       \\
     &           &                   &        &            &       &  1.04 &  0.87 &              &        &       \\
[1 ex]
  38 &         6 & J20562218+443720  &   83.5 & 20:56:22.1 & 12.59 &  0.52 &  0.44 &              & A3 V   &  $lm$ \\
     &           &                   &  -40.3 & 44:37:20.0 &       &  0.63 &  0.55 &              &        &       \\
     &           &                   &        &            &       &  1.06 &  0.97 &              &        &       \\
[1 ex]
  48 &        13 & J20562809+443501  &   18.7 & 20:56:28.1 & 13.19 &  0.80 &  0.75 &              & A2 V   &  $nm$ \\
     &           &                   & -180.9 & 44:34:58.9 &       &  0.43 &  0.38 &              &        &       \\
     &           &                   &        &            &       &  0.96 &       &              &        &       \\
[1 ex]
  53 &        28 & J20564026+443926  & -109.2 & 20:56:40.2 & 13.44 &  0.79 &  0.71 &              & A3 V   &  $pm$ \\
     &           &                   &   86.6 & 44:39:26.1 &       &  0.66 &  0.58 &              &        &       \\
     &           &                   &        &            &       &  1.25 &  1.16 &              &        &       \\
\hline
\end{tabular}
\begin{tabular}{c}
\begin{minipage}{16cm}
\vspace{0.1cm}
{\bf Notes:} \\
- \# and \#$_{ZS}$ columns indicate our numbering and that from 
Zdanavicius \& Straizys (1990), respectively. \\
- \#$_{ZS}$ numbers with an asterisk indicate stars separated in two components 
(values correspond to the brightest one) \\
- The full Table~2 is available, with all the photometric measurements, 
in electronic version at the CDS. \\
\end{minipage}
\end{tabular}
\end{center}
\end{table*}

\subsection{Proper motions}

$\hspace{0.5cm}$
Important information on the kinematics and membership of the brightest stars in and 
around a star cluster might in principle  be derived from the proper motions as 
available in the Tycho-2 catalogue (H{\o}g et al. 2000). With this aim in mind, we 
collected proper 
motion components for 22 stars in a field of $10^{\prime}$ radius centered in 
NGC~6996. They are shown as a vector point diagram in Fig.~6. The points 
distribution is characterized by a clump of several stars and few others 
placed around it. This fact can be readily interpreted as 
indicative of the presence of a star cluster.
However, on the basis of the analysis here below, we found only 3 member stars
with available proper motion from Tycho-2, and therefore we restrain from
any kinematic analysis of NGC~6996 stars.\\

\subsection{Photometric diagrams}

$\hspace{0.5cm}$
The color-color diagrams (CCDs) and the color-magnitude diagrams (CMDs) from our 
data are shown in Figs.~7 and 8, respectively. In Fig.~9 we present the 
corresponding CMDs from 2MASS data for the cluster region stars 
($R < 7^{\prime}$) and for a region around it (see caption) that is used as  
{\em comparison~field}. Some remarkable features in all these diagrams are the following 
ones: 

\begin{itemize}
\item The Main Sequence (MS) in Fig. 7, 8 and 9 is significantly wide, a fact that we
      mainly ascribe to differential reddening due to the patchy distribution
      of gas and dust inside the cluster itself;
\item the reddening (see Fig.~7b) law inside NGC~6996 does not seem normal, and
      turns out to be significantly different from that commonly accepted to
      hold for the Cygnus region (Johnson 1965). Indeed
      most of the stars follow a parallel path to a higher 
      excess ratio.  This is quite an usual situation 
      in very young open clusters and star forming regions (e.g. V\'azquez 
      et al. 1996, Carraro et al. 2003) but not so common in older clusters;
\item a noticeable gap appears at $B-V \approx 0.6$ and $V \approx 13$. This is
      not an unexpected feature, since several other young or intermediate age
      clusters have been found to exhibit gaps (e.g. Giorgi et al. 2002; 
      Yadav \& Sagar 2002; Baume et al. 2003b). \\
\end{itemize}

\begin{figure}
\centering
\includegraphics[width=9.cm]{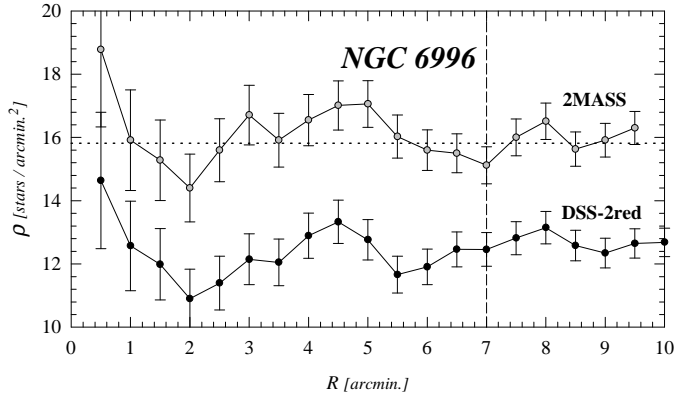}
\caption{{} Stellar density in the region of NGC 6996 as a function of the
radius from 2MASS and DSS-2 red data. The dashed line indicates the adopted limit 
for the cluster.}
\end{figure}

\begin{figure}
\centering
\includegraphics[width=7.5cm]{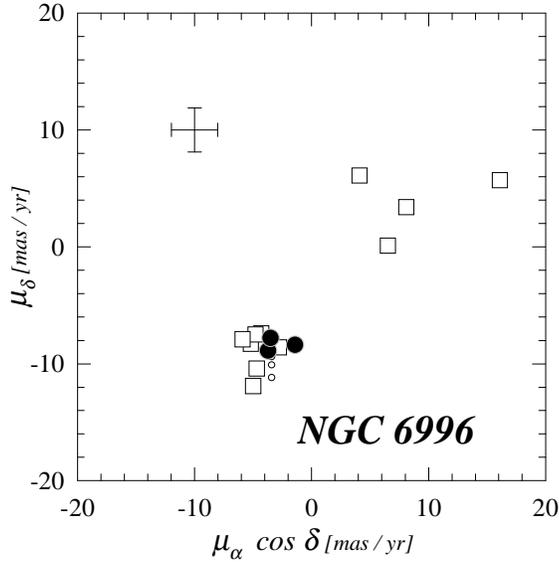}
\caption{{}Vector point diagram for the stars $10^{\prime}$ around
the center of NGC 6996 (data from Tycho-2 catalogue). See Fig. 7 caption for 
the meaning of the symbols. The cross  indicates the mean error values.}
\end{figure}

\subsection{Cluster membership}

$\hspace{0.5cm}$
As a first step, we base our membership assignment procedure
on a synoptic analysis of the star 
positions in  the various photometric diagrams (e.g. Baume 1999, 2003ab; Carraro 2002). 
By inspecting Fig.~7a, we notice  that up to $(B-V) \approx 0.6$ 
stars are placed onto an apparent well recognizable MS composed by 
B and later type stars according to a Schmidt-Kaler's (1982) ZAMS. They have also 
a compatible position on the CMDs of Fig.~8 and 9 down to $V \approx 13.5$. 
However, the considered B-type stars are more dispersed in these diagrams.\\

As a second step, we combine the magnitudes and colors with our spectral 
classification and that given by Zdanavicius \& Straizys (1990), when available, 
and the result is that most of the brighter stars have very low excess values 
or/and low distance modulus, and only A-type stars have acceptable solutions. 
Therefore, the stars from the first group were considered as cluster non members 
($nm$, probably interlopers), whereas the later ones as 
likely cluster members ($lm$). Additionally, 
following the ZAMS path along the CCDs and CMDs toward later than A spectral types  
we identify other stars with compatible positions on the photometric 
diagrams. Here however the stellar contamination by field stars starts to 
become important, and  therefore we adopt them as probable cluster members ($pm$). This 
membership assignment produces an amount of stars in  different bins of $V$ 
magnitude that is in agreement with the over-excess of stars present in Fig.~9a 
when compared to Fig.~9b. \\

Stars $\#_{ZS}$ 6 and 14 (numbering from Zdanavicius \& Straizys 1990) deserve 
special attention. The former lies below the ZAMS in the CCD of Fig.~7a, and
by inspecting also its position in the CMDs we interpret its color as due
to binarity, the secondary being a cool red star.\\
The latter is a red star classified as
a K2 III. Looking at  its location in  the CMDs, also in relation with the superposed 
isochrones (see below),  we are inclined to consider it a cluster member. Unfortunately, it
has no proper motion measurements (from Tycho-2) and therefore we adopt it as
a probable member ($pm$). \\

\begin{table} 
\fontsize{8} {10pt}\selectfont
\tabcolsep 0.50truecm 
\caption{Computed $R_V$ values by using different methods.} 
\begin{tabular}{ll} 
\hline 
${\bf Method}$                  & Obtained $R_V$ value \\
\hline
{\it Variable Extinction} & 3.4 $\pm$ 0.2 \\
{\it Excesses Relation}   & 4.2 $\pm$ 0.1 \\
{\it Color Difference}    & 4.0 $\pm$ 0.4 ($s.d.$) \\
\hline
\end{tabular}
\end{table}

\begin{figure*}
\centering
\includegraphics[height=10cm]{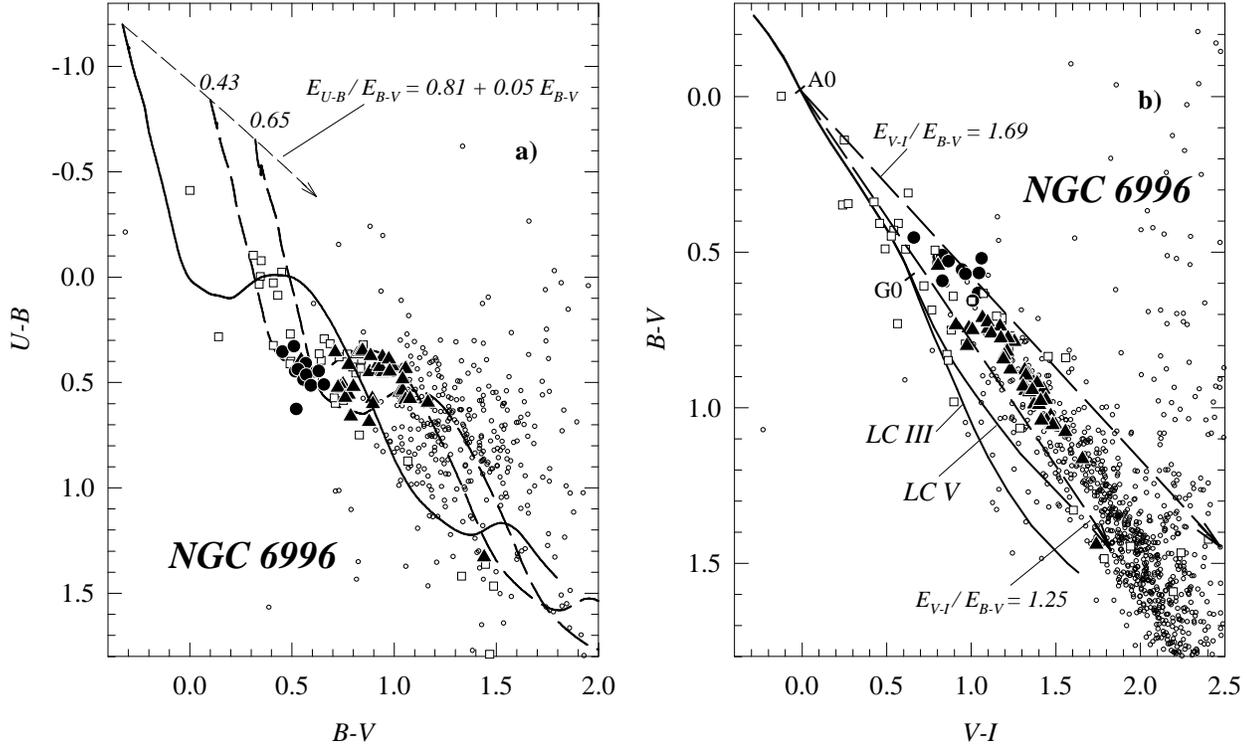}
\caption{{} Color-color diagrams (CCDs) of stars in the region of NGC 6996. 
{\bf a)} $U-B$ vs. $B-V$ diagram. Symbols have the following meaning: circles are 
likely member stars ($lm$), triangles are probable member stars ($pm$), white 
squares are non members ($nm$) and small hollow circles are stars without 
any membership assignment. The solid line is the Schmidt-Kaler's (1982) ZAMS, whereas 
the dashed lines are the same ZAMS, but shifted by $E_{B-V} = 0.43$ and $0.65$, 
respectively ({\bf see also Fig~10a}). The dashed arrow indicates the reddening path. 
{\bf b)} $B-V$ vs. $V-I$ diagram. Symbols as in Fig. 7a. Solid lines are the 
intrinsic positions for stars of luminosity classes V and III (Cousins 1978ab). 
The two dashed lines give the typical excess ratio for the Cygnus region ($E_{V-I} /
E_{B-V} = 1.25$) and that adopted for the cluster ($E_{V-I} / E_{B-V} = 1.69$).}
\end{figure*}

\begin{figure*}
\centering
\includegraphics[height=10cm]{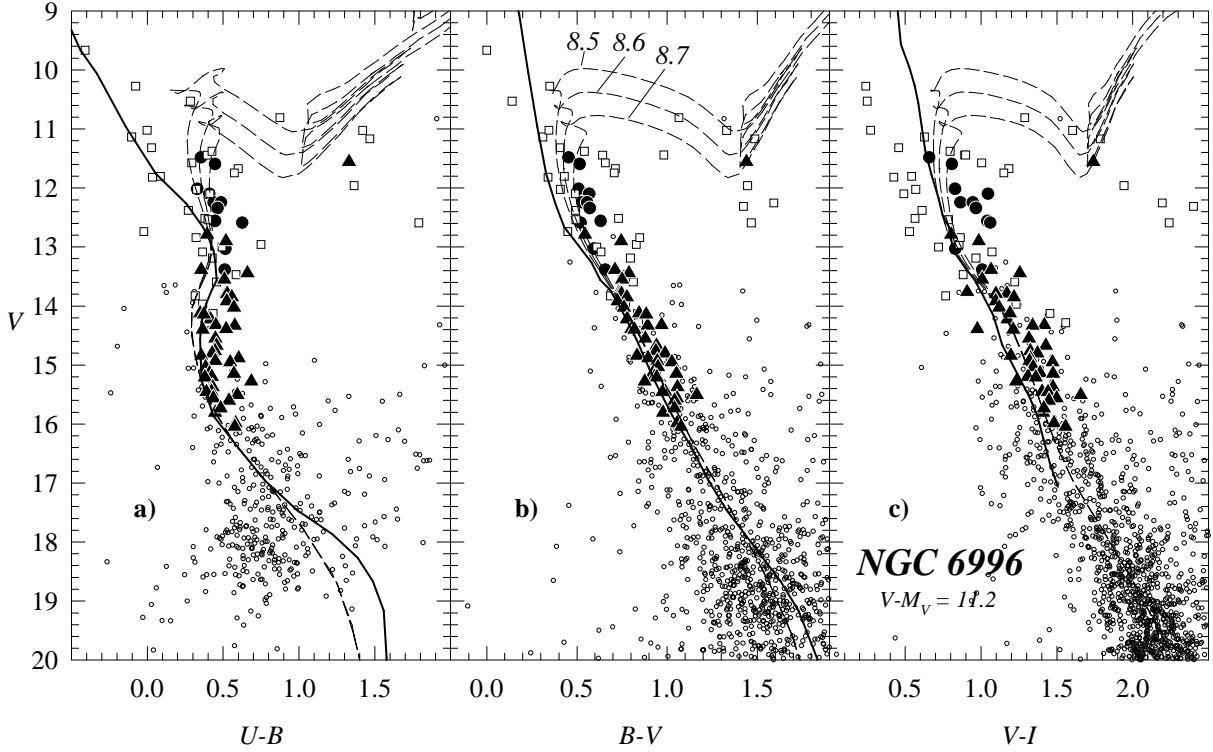}
\caption{{} Color-magnitude diagrams (CMDs) for all the stars covered in the 
field of NGC 6996. Symbols as in Fig. 7a. The solid line is the Schmidt-Kaler 
(1982) empirical ZAMS shifted by  the apparent distance modulus $V-M_{V} = 
11.2$ ($V-M_{V} = V_{0}-M_{V} + 4.1 \times (E_{B-V})$, see section 4).The dashed curves are 
the isochrones from Girardi et al. (2000). The reported numbers give 
the $\log(age)$.}
\end{figure*}

\begin{figure*}
\centering
\includegraphics[height=10cm]{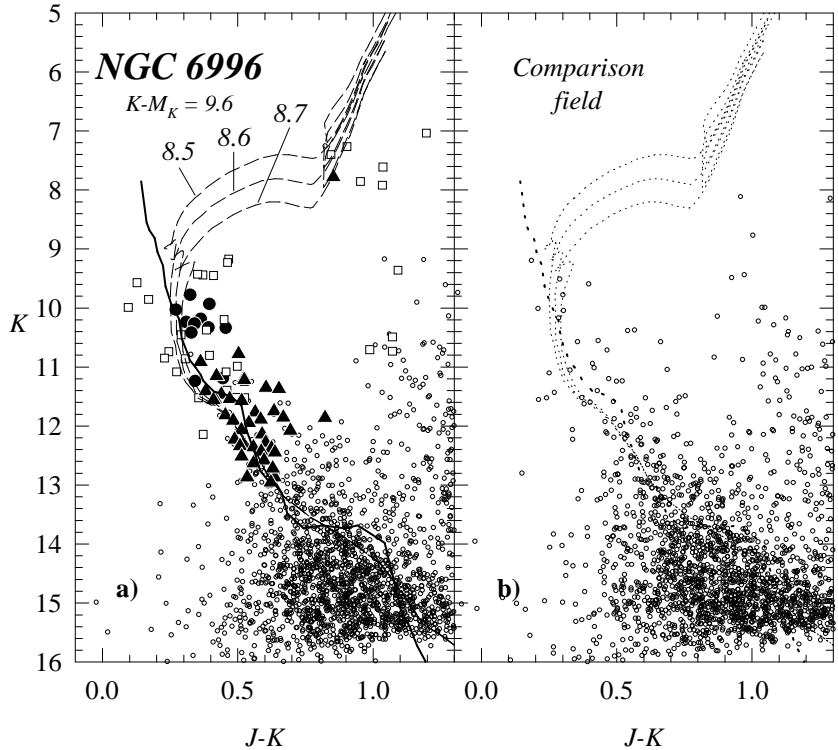}
\caption{{} CMDs from 2MASS catalog. Symbols as in Fig. 7a. {\bf a)} Stars 
placed inside the cluster area ($R < 7^{\prime}$, see Fig.~5). {\bf b)} Stars 
located in the comparison field around the cluster ($R > 8\farcm85$ and inside 
a box $20^{\prime} \times 20^{\prime}$). The solid line in panel {\bf a)} and 
the dotted one in panel {\bf b)} are the intrinsic position for MS stars from 
the Schmidt-Kaler (1982) and Koorneef (1983) calibrations fitted to the apparent 
distance modulus $K-M_K = 9.6$ ($K-M_K = V_{0}-M_{V} + 
(3.1-2.78)~(4.1/3.1)~E_{B-V}$, see section 4). The dashed curves are 
the isochrones from Girardi et al. (2000). The reported numbers give 
the $\log(age)$.}
\end{figure*}

\begin{figure}
\centering
\includegraphics[width=7.5cm,height=16cm]{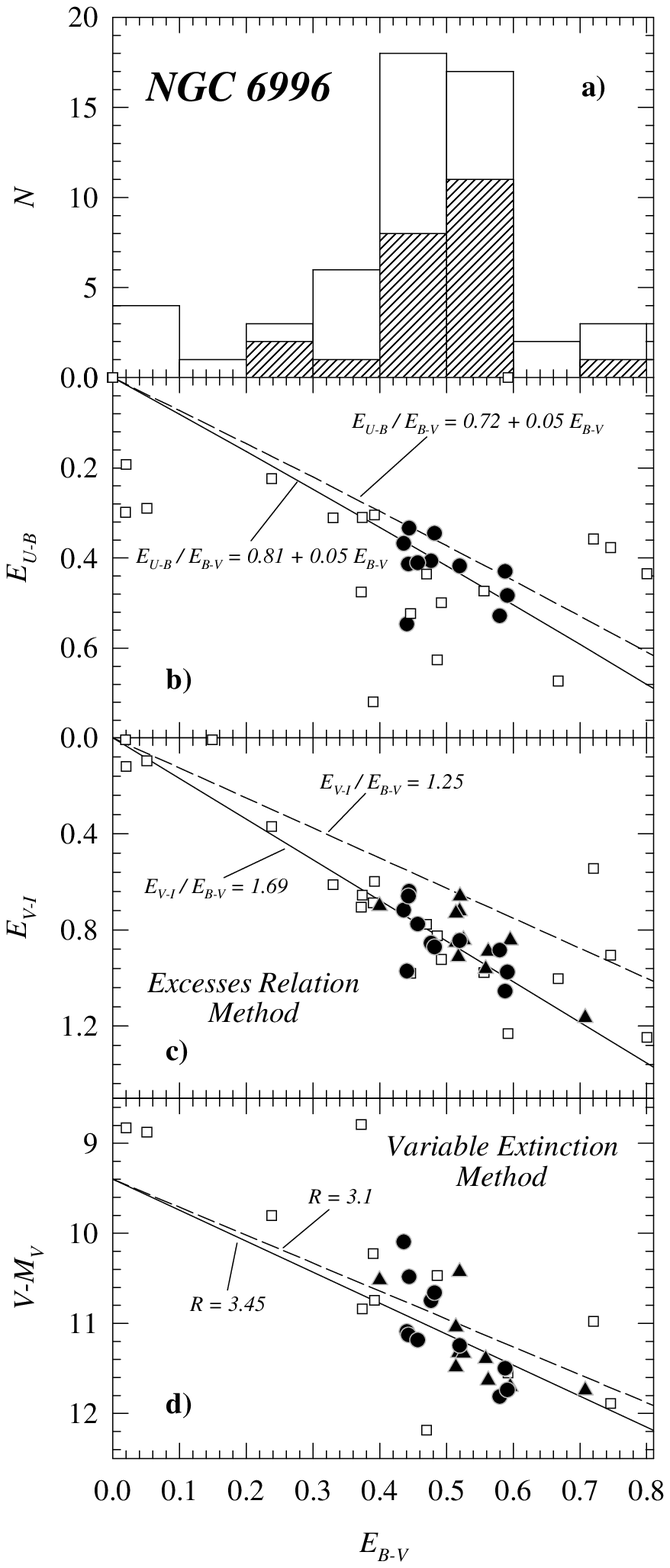}
\caption{{} {\bf a)} $E_{B-V}$ distribution for stars inside the $7^{\prime}$ 
cluster radius (white histogram) and for adopted likely and probable cluster 
members (dashed histogram). {\bf b)} $E_{U-B}$ vs. $E_{B-V}$ diagram for stars 
with available spectral classification. {\bf c)} $E_{V-I}$ vs. $E_{B-V}$ 
diagram. {\bf d)} $Variable~extinction~method$. Symbols in panels {\bf b)}, 
{\bf c)} and {\bf d)} as in Fig. 7a. Dashed lines represent normal relations 
and solid ones those obtained for NGC 6996.}
\end{figure}

\section{Cluster basic parameters}

\subsection{Corrected colors and magnitudes}
$\hspace{0.5cm}$
In order to infer the intrinsic colors, when spectral classification is available we
derive individual excesses by  using Schmidt-Kaler (1982) relations. Figs.~ 7a  and
10b 
show that the typical relation $E_{U-B}/E_{B-V} = 0.81 + 0.05~E_{B-V}$ for the 
Cygnus region (Johnson 1965) fit the point distribution much better than 
the normal one ($E_{U-B}/E_{B-V} = 0.72$). We use then the former relation to 
obtain some additional individual excesses from the CCD of Fig.~7a for stars 
without spectral classification. Then by considering all the adopted likely members and probable 
members with $V < 14$, we found a mean excesses of $E_{B-V} = 0.52 \pm 0.08~(s.d.)$ 
and $E_{U-B} = 0.44 \pm 0.08~(s.d.)$. 
The mean values are then adopted as representative of the 
cluster color excess, whilst the lowest value is interpreted as the foreground 
color excess. \\

As for the reddening law in the direction of NGC~6996, it readily appears to
be anomalous, and therefore we proceeded  to compute the ratio $R_V = A_V/E_{B-V}$ 
by using the {\it Excesses Relation Method} (Fig.~10c), the {\it Variable Extinction 
Method} (Fig.~10d) and the {\it Color Difference Method} (Fig.~11). 
To build  
Figs.~10c and 10d up, we compute individual $E_{B-V}$, $E_{V-I}$ and $V-M_V$ values by means of 
the Schmidt-Kaler (1982) and Cousins (1978ab) calibration relations, a method 
also applied in Tr 14 (V\'azquez et al. 1996) and in NGC 3293 (Baume et al. 2003). 
In the case of the {\it Color Difference Method} we combined the information from 
spectral classification, optical photometry and near infrared one
from the 2MASS. For other applications of this method, see Th\'e \& Graafland (1995)
and Carraro et al. (2003).\\

It is clear from Fig.~10c that star positions follow a path different from the 
typical one for Cygnus region ($E_{V-I}/E_{B-V} = 1.25$) resulting in a higher 
$R_V$ value. This is not very evident in Fig.~10d where there is a larger spread due 
probably to the presence of binaries. However, by performing least squares fittings 
over Figs.~10c and 10d and an extrapolation in Fig.~11, we obtain almost similar 
values (see Table~3). Therefore we adopt $R_V = 4.1$ to compute corrected 
magnitudes $V_0 = V - R_V \times E_{B-V}$ for cluster likely and probable members. This 
$R_V$ value yields a mean cluster absorption  $\langle A_V \rangle = 2.13$. 
It is worth noticing that this estimate  is 
comparable with the absorption map in this region obtained by Goudis \& White 
(1979). \\

\subsection{Cluster distance and age}

$\hspace{0.5cm}$
The distance of NGC~6996 is derived by superposing the Schmidt-Kaler (1982) ZAMS 
onto the reddening-free CMD (Fig.~12). The best ZAMS fitting was achieved for a 
distance modulus $V_0-M_V = 9.4 \pm 0.2$ (error from inspection). We also apply 
the spectroscopic parallax method to 10 likely and probable member stars of 
luminosity class V (see Table~3) by using the relation of spectral types and 
$M_V$ from Schmidt-Kaler (1982). This method yields a value $V_0-M_V = 9.0 \pm 
0.4~(s.d.)$. The large dispersion of the last value and its difference from the
ZAMS fit can be ascribed to the presence of binary and somewhat evolved stars included in 
the computation. The adopted distance modulus is therefore $V_0-M_V = 9.4 \pm 
0.2$ which in turn implies that  NGC~6996 is located $760 \pm 70$ pc away from the Sun. \\

As for the age of NGC~6996, we over-imposed on the CMD (see Fig.~12) 
a set of isochrones derived from Girardi et al. (2000)
evolutionary models (computed with solar metallicity, mass loss and overshooting).
The fit has been performed taking into account only MS stars
and is compatible with an age for the cluster of about 
310 Myr ($\log(8.5)$). By assuming that  stars having A0 spectral type are still along the 
MS, we derive again an age near to 390 Myr ($\log(8.6)$) for this 
cluster. Both procedures yield then similar results and we 
adopt $350 \pm 50$ Myr as the cluster age. \\

\section{Discussion and conclusion}
$\hspace{0.5cm}$
We have presented the first multicolor CCD photometric study 
in the region of the open cluster 
NGC~6996 together with spectral classification of some bright stars. 
NGC~6996 turns out to be  a moderate age open cluster ($\approx 350$ Myr) 
located close to  the west edge of NGC~7000, the North America Nebula. 
Our analysis places the cluster at a distance of $760 \pm 70$ pc. 
The evidence emerges of a significant differential reddening affecting the stars' 
positions in  all the photometric diagrams. 
We also point out that the reddening law has an anomalous 
value in the direction of this cluster.\\

\noindent
To address the issue of its possible connection with the Nebula,
we firstly summarize what it is known about the distance of the Nebula itself
(see for reference Fig.~1).
According to Straizys et al. (1999)
the dust cloud that separates the North America and Pelican nebulae is placed at
about 580 pc form the Sun. As for H II region (NGC~7000 itself), the distance is
far from being reasonably contrained. In fact 
different works claim for very different distance 
estimates, ranging from 420 pc (Beer 1964) to 1980 pc (Dieter 1967).
HD~199579, an O6 star considered responsible for at least part of the excitation
of the region, is placed at a distance of 1200 pc by Miller (1968) and 830 pc
by Garmany \& Stencel (1992), which consider the star a member
of the Cyg OB7 association. According to many other authors 
however, the most accepted value seems to be  $\sim 1$ kpc 
(Downes \& Rinehart 1966; Wendeker 1968;  Goudis 1976; Bally 
\& Scoville 1980). \\
If we accept this value as the distance of NGC~7000,
NGC~6996 turns out to have no apparent relationship with the HII region,
the cluster being placed about 300 pc closer to the Sun.
The cluster age, and the absence of early spectral type stars further
corroborate this hypotesis.\\

\noindent
The computed $R_V$ value for the cluster region is higher than the common one
holding for the Galaxy (3.1, Mathis 1990), 
and at odd with previous studies (e.g. Cambr\'{e}sy et al. 2002) which assign to  NGC~7000
a normal $R_V$ ratio. 
High $R_V$ values are indicative of the presence of dust grain of
large size, typically larger than 0.05 micron. Since 
NGC~6996 is dominated by A type stars, UV radiation is not very effective, and
one expects that dust grains grow in size (Kim \& Martin 1996).
Alternatively, another possibility would be that small size dust grains
have been kicked off the cluster by a presumed population of massive stars
already died as type II SN\ae~ (McKee 1989), a scenario which is quit compatible with
the age of the cluster.

\begin{figure*}
\centering
\includegraphics[width=13cm]{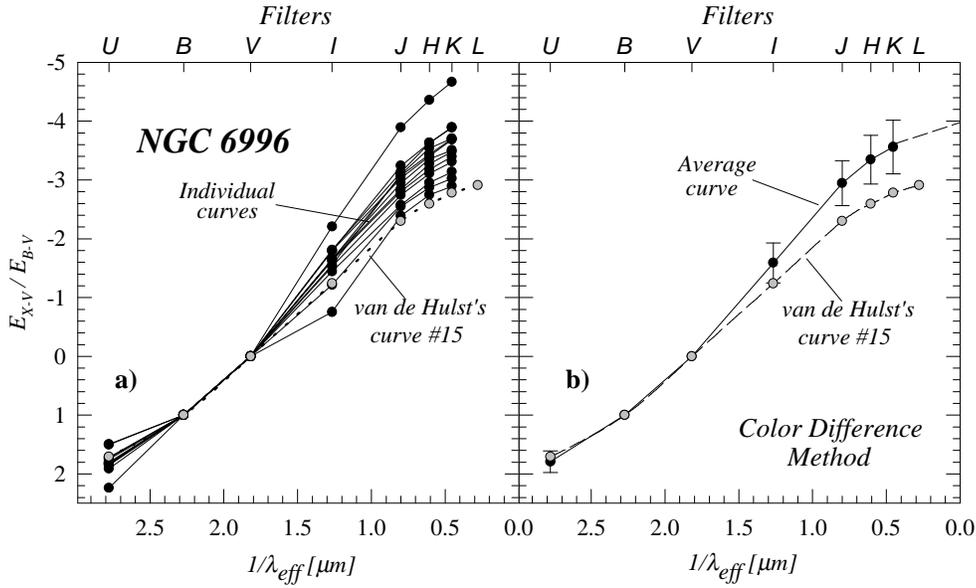}
\caption{{} The color difference method. {\bf a)} Solid lines are individual 
curves computed for stars with spectral classification and luminosity class V 
(see Table~2) {\bf b)} Solid line is the obtained average curve together 
with their extrapolation up to $1/\lambda_{eff} = 0.$. Dotted lines with grey 
simbols in both panels are the extinction curve \#~15 from van de Hulst 
(Johnson 1968).}
\end{figure*}

\begin{figure*}
\centering
\includegraphics[height=10cm]{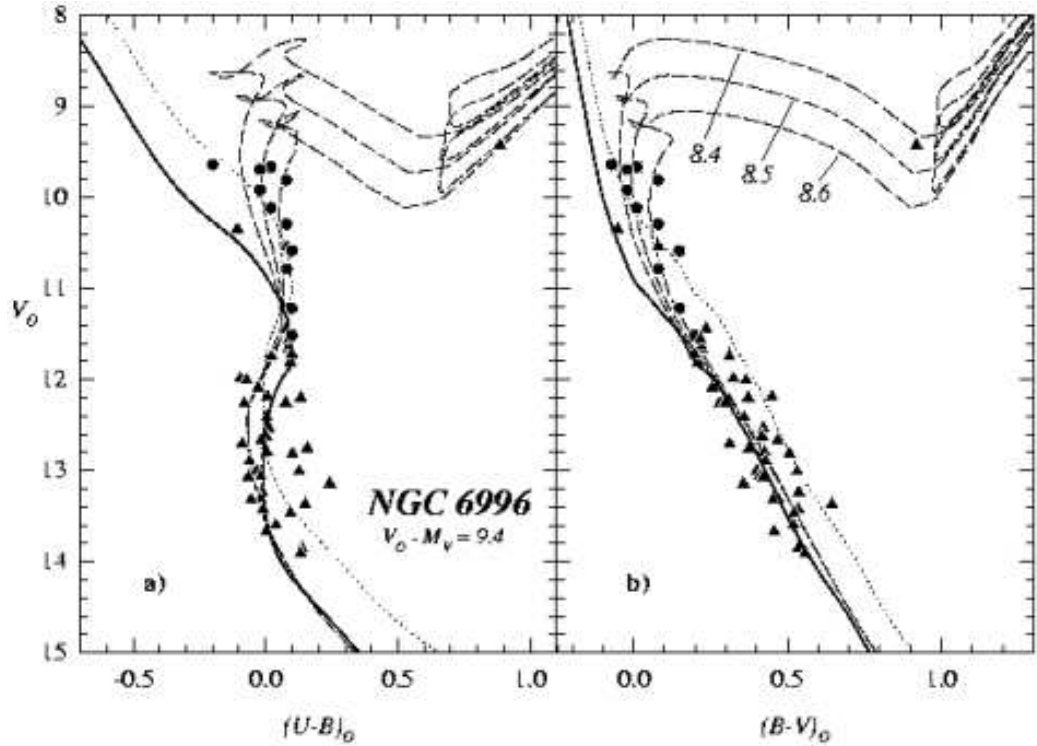}
\caption{{} $V_O$ vs. $(B-V)_0$ diagram of the likely and probable member 
stars in the region of NGC~6996. Symbols as in Fig. 7a. Solid and dotted lines 
are the Schmidt-Kaler (1982) ZAMS shifted by the adopted distance modulus 
$V_O - M_V = 9.4$ and the $0.75$ envelope limit for binaries. Dashed curves are 
the isochrones from Girardi et al. (2000). Numbers give the $\log(age)$.}
\end{figure*}

\begin{acknowledgements}
The authors acknowledge the Asiago Observatory staff for the technical 
support and the director of Teramo Observatory for the generous
time allocation. The work of G.B. is supported by the Universit\`a di Padova
(Italy) through a postdoctoral grant. 
\end{acknowledgements}

\end{document}